# Time-resolved monitoring of polycyclic aromatic hydrocarbons adsorbed on atmospheric particles

Gustavo Sousa[1], Denis Kiselev[1], Jérôme Kasparian[1,*], Christian George[3], José Ferreira[2], Philippe Favreau[2], Benoît Lazzarotto[2], Jean-Pierre Wolf[1]

1. Université de Genève, GAP, Chemin de Pinchat 22, CH-1211 Geneva 4, Switzerland

2. État de Genève – DETA – SABRA, Avenue de Sainte-Clotilde 23, CP 78, CH-1211 Geneva 8, Switzerland

3. Univ Lyon, Université Claude Bernard Lyon 1, CNRS, IRCELYON, F-69626, Villeurbanne, France

* e-mail: jerome.kasparian@unige.ch; telephone: +41 (0) 22 379 05 12; fax: +41 (0) 22 379 05 59

**Acknowledgments**

We acknowledge unvaluable experimental assistance from Francesco Battaglia, Elicio Délicado, and Pierre-Emmanuel Huguenot (État de Genève – DETA – SABRA).

We also acknowledge funding by the Swiss National Science Foundation through the NCCR MUST (Molecular Ultrafast Science and Technology) Network. J.P. Wolf acknowledges support from the European Research Council ERC-2013-PoC, Grant 632156, « LIPBA ».

**ABSTRACT:** Real-time monitoring of individual particles from atmospheric aerosols was performed by means of a specifically developed single-particle fluorescence spectrometer (SPFS). The observed fluorescence was assigned to particles bearing polycyclic aromatic hydrocarbons (PAH). This assignment was supported by an intercomparison with classical speciation on filters followed by gas chromatography-mass spectrometry (GC-MS) analysis. As compared with daily-averaged data, our time resolved approach provided information about the physicochemical dynamics of the particles. In particular, distinctions were made between background emissions related to heating, and traffic peaks during rush hours. Also, the evolution of the peak fluorescence wavelength provided an indication of the aging of the particles during the day.

# Introduction

In the recent years, air pollution by particulate matter (PM) has become a major health issue in large urban areas (Pope and Dockery 2006; Ito et al. 1995). In particular combustion-related carbonaceous particles, which contain Polycyclic Aromatic Hydrocarbons (PAH), are recognized as hazardous for human health since they increase the risk of respiratory tract cancers. Since 2004, the EU-Commission has fixed a target value for the benzo(a)pyrene concentration in air at 1 ng/m$^3$ for the total content in PM$_{10}$ (particulate matter with aerodynamic diameter up to 10 μm) averaged over a calendar year (2004/107/EC (EU-Commission 2004) and 2015/1480 (Commission directive 2015)). EU-directives also identify a series of similar PAH species that should be monitored, including benzo(a)anthracene, benzo(b)fluoranthene, benzo(j)fluoranthene, benzo(k)fluoranthene, indeno(1,2,3-cd)pyrene, and dibenz(a,h)anthracene.

The official monitoring method according to the EU is described in the directives 2004/107/EC (EU-Commission 2004) and 2015/1480 (Commission directive 2015). Taking into account these directives, PM$_{10}$ are collected on filters during 24 hours and a composite sample representing a month period is subsequently analyzed in the laboratory by means of chromatography coupled to mass spectrometric detection. However, these methods are expensive and time consuming, so that permanent monitoring with a higher time resolution is often impossible.

Still, considering the fast dynamics of the atmosphere, real-time measurements would be highly valuable for identifying emission sources (for instance traffic *vs*. heating), as well as to characterise actual exposure levels. For this reason, substantial effort has been dedicated to the development of optical counters, able to identify PAH-containing particles using fluorescence detection (Pinnick et al. 2004; Pinnick et al. 2013; Robinson et al. 2013; Pan 2015; Miyakawa et al. 2015). The most advanced methods use spectrally resolved fluorescence measurements of individually illuminated particles, which allows sorting the signatures into « clusters » featuring similar spectra (Pan et al. 2003; Pinnick et al. 2004; Pinnick et al. 2013). These signatures may in principle be subsequently attributed to specific fluorophore mixtures.

In the present paper, we report the results of an almost 1 month long campaign during winter 2013-2014 at an urban location of Geneva, Switzerland, near to moderated-traffic roads, using a dedicated, specifically developed single-particle fluorescence spectrometer (SPFS) (Kiselev et al. 2013). We checked that the daily-averaged concentration of PAH-bearing particles from the SPFS was consistent with the PAH load in the atmosphere as measured by filter sampling. Furthermore, the high temporal resolution offered by the identification of individual particle could provide access to the dynamics of both emission and aging of particle-adsorbed PAH.

# Materials and methods

PAH contained in $PM_{10}$ were investigated in an urban location of Geneva at the "Sainte Clotilde" station of the Geneva state air quality monitoring network (ROPAG), 7 m away from a street with moderate traffic (estimated ~5'000 vehicles/day) and 3 m above ground. The instruments themselves were located in a shelter that protected them from precipitation but kept them close to the outside temperature. The measurement station was also equipped with particle analyzers (high volume sampler - Digitel DA80; optical particle analyser - Grimm EDM #180). Measurements took place from 27/11/2013 to 19/12/2013, for a total of 532 hours. Table 1 summarizes the main specifications of the instruments that have been deployed simultaneously.

The GAP-SPFS (Fig. 1) (named in analogy to Pinnick et al., 2013) recorded both the scattering intensity and the full visible fluorescence spectrum of each individual particle in real time. As described in Kasparian et al. (2017), it measured the scattering and the fluorescence spectrum of individual aerosol particles in a concentrated air flow. In order to achieve this, the air was sampled at 60 L/min, concentrated by a 10x aerodynamic lens, and then focused by a sheath nozzle into a stream of 500 µm diameter (Kiselev et al., 2013). This stream crossed infrared laser beams for the scattering measurements, as well as a UV laser beam for exciting the fluorescence (Hill et al. 1999 and Pan et al. 2001) that was detected by a multi-angle laser scattering module (Bonacina et al. 2013). This procedure also allowed measuring simultaneously the aerodynamic size of each particle. Active "on the fly" modulation of the UV laser intensity was also applied to prevent saturation, so that particles between 1 µm and 60 µm diameter could be analyzed. Scattering signals were also used to trigger the pulsed UV laser (337 nm), which excited the fluorescence of the same individual particle. The fluorescence was retrieved by a reflective lens and analyzed in the 360–650 nm spectral region with a 32-channel spectrometer based on a multi-anode photomultiplier (Hamamatsu H7260-03).

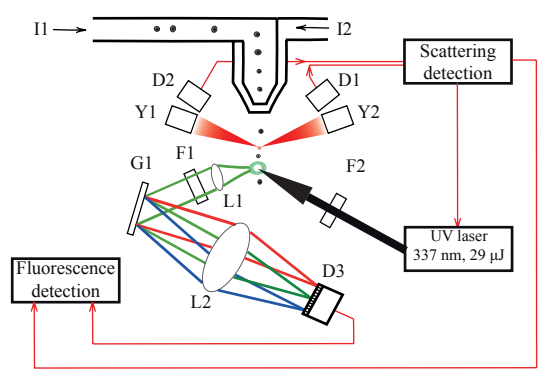

**Fig. 1** Principle of the Single-particle fluorescence spectrometer. I1, I2: Air inlet; Y1, Y2: Infrared laser diodes; D1, D2: Photomultipliers; L1: Reflective objective; L2: Bi-convex lens; F1, F2: Filters; G1: Diffraction grating; D3: 32-anode photomultiplier

**Table 1** Instruments and data available during the measurement campaign.

| Device | Measured parameters | Remarks |
| --- | --- | --- |
| Optical particle analyzer (Grimm EDM #180) | $PM_{10}$ mass concentration | Continuous measurement (6 s resolution), 1/2 h average |
| $PM_{10}$ exposed filters (24h collection with a Digitel high volume sampler DA80) | Mass concentration in air of 16 individual PAH species quantified by GC-MS analysis in $PM_{10}$ collected on filters | Daily accumulation |
| Single-particle fluorescence spectrometer ("GAP-SPFS") (Pan et al. 2001) | Particle concentration and UV fluorescence spectra of individual particles between 1 μm and 60 μm diameter | Continuous measurement, hourly average |

We detected and sorted the individual particles according to their spectrum through a self-referencing procedure. First, fluorescent particles were selected among all particles, based on a threshold in the total fluorescence spectrum, ensuring an adequate signal-to-noise ratio to allow spectrum analysis. Then, the particle spectra were smoothed using a third-order polynomial and normalized to a peak value of 1. If the particle spectrum was similar to a previously detected one, it was assigned to the same cluster. If not, this spectrum was used as the prototype to define a new cluster. The procedure was then applied a second time after averaging all spectra within each cluster, so as to avoid over-weighting the particle that initially defined the cluster prototype. In this procedure, spectra were considered similar if (i) their intensity had a correlation coefficient over 0.85, and (ii) the relative difference between their respective spectrally-averaged intensities was less than 30%. As the spectra are normalized to a peak value of 1, this latter criterion distinguishes between spectra displaying sharp peaks and those with a smooth maximum over a wide background. This procedure allowed us to identify 15 particle clusters accounting for 1% or more of the fluorescent particles, with some overlap. The thresholds of 0.85 and 0.3 on correlation and relative intensity differences have been chosen to keep a reasonable number of families while limiting the overlap between them, i.e., the risk that a given spectrum is simultaneously assigned to two or more families.

In addition, the average particle fluorescence spectra varied during the day, evidencing an overall evolution of the population of fluorescent particles. To characterize this evolution, we considered the wavelength $\lambda_{max}$ of the maximum of each particle fluorescence spectrum. The average of $\lambda_{max}$ over all particles during the considered time interval was taken as the characteristic wavelength of the particle population.

Over the same time period, $PM_{10}$ were collected daily on glass fiber filters using a high volume sampler. Before and after the exposition, the filters were weighed at fixed temperature and relative humidity. Taking into account the volume of ambient air probed, the daily concentration of $PM_{10}$ could be calculated. The exposed filters were kept at 4°C until analysis. They were subsequently prepared for the characterization of the Environmental Protection Agency's 16 priority pollutant PAH. Each filter was spiked with $^{13}C_4$-benzo(a)pyrene that was used as an internal standard and extracted with dichloromethane by sonication. The extract was then concentrated and analyzed by gas chromatography-mass spectrometry (GC-MS) following a validated operating procedure.

Results were obtained for all individual compounds and day-averaged to get the total particulate PAH content in ng/m$^3$.

## Meteorology and air pollution during the measurements

The following meteorology summary is mainly based on MeteoSwiss data and reports. The entire period was mainly dry - few rainfalls occurred on the 30$^{th}$ of November and on the 14$^{th}$ of December, 2013 - and the temperature was between –5°C and +5°C. Two significant periods of temperature inversion occurred from the 01$^{st}$ to the 05$^{th}$ of December, with some north wind at the beginning and then from the 09$^{th}$ to the 17$^{th}$ of December with a high-pressure situation and no wind. The PM$_{10}$ daily maximum concentrations measured at the ROPAG's Sainte-Clotilde station during these two inversion periods were 62 μg/m$^3$ on the 4$^{th}$ of December and 72 μg/m$^3$ on the 11$^{th}$ of December.

## Results and discussion

A total of 5.449.801 particles were detected by the SPFS, corresponding to an average of 10.244 particles per hour. Among them, ~2% were fluorescent. Fig 2 displays the spectra of the 10 most abundant clusters. Due to the wide variety of emitted PAH species, each individual particles carried a cocktail of species rather than a single one. The observed fluorescence spectrum was therefore a weighted average of all adsorbed molecules. Furthermore, no reference spectrum is available in the literature for adsorbed PAH, since published reference spectra have been measured in cyclohexane (Finlayson-Pitts and Pitts, 2000). For these two reasons, a direct identification in terms of individual PAH species was impossible. Still, their peaks between 450 and 550 nm corresponded to PAH of various sizes, from bicyclic to polycyclic molecules (Hayashida et al. 2006). In particular, the clusters peaking around 450 and 500 displayed similarities with fluoranthene and pyrene and with benzo(a)pyrene, respectively (Pöhlker et al. 2012). The 550 nm band corresponds to fulvic acid, although the latter does not belong to PAH (Pöhlker et al. 2012). The extension of spectra up to the green spectral region may be due to contributions from charge-transfer complexes (Phillips and Smith, 2015). It should be noted that none of these families can be identified with bacterial fluorescent spectra as measured by Saari et al. (2013).

Interference with biological aerosols (Pan et al. 2001; Pan et al. 2003; Kiselev et al. 2013; Pinnick et al. 2013) or biological material adsorbed on mineral particles (Maki et al. 2008) that can be transported by wind (Miyakawa et al. 2015) may have occurred. However, the density of fluorescent particles (See Fig. **5**) changes quickly. As the measurements were recorded in a street canyon far from biogenic emitters that would induce quick changes (Huffman et al. 2013), such fast dynamics call for attributing them mainly to anthropogenic sources rather than to biological particles. This is also consistent with the finding by Herrmann et al. (2006) that biogenic aerosols can be neglected in an European winter urban environment, and with the immediate proximity of the sampling site to road traffic and domestic heating sources, as well as the low level of biological material (especially pollens) in winter. Therefore, in the following, we considered the fluorescent particles as mostly anthropogenic PAH-bearing particles.

The relative abundance of the different clusters, hence of the different particle types, stayed quite constant along the whole campaign. Indeed, pairwise correlations of the temporal series of all clusters and of the total of fluorescent (i.e., PAH-bearing) particles, showed very significant positive correlations coefficients (mostly above 0.9). In the following we therefore focused on the count of fluorescent particles, rather than on individual clusters. However, the cluster spectra displayed on Fig 2 allows comparison with previous data acquired by the ARL-Yale group in US cities (Pan et al. 2012). It is interesting to observe some very similar fluorescence shapes (like Clusters 1, 3, and 6), although both the cluster identification algorithm and the automotive park and gasoline/diesel cocktails were different.

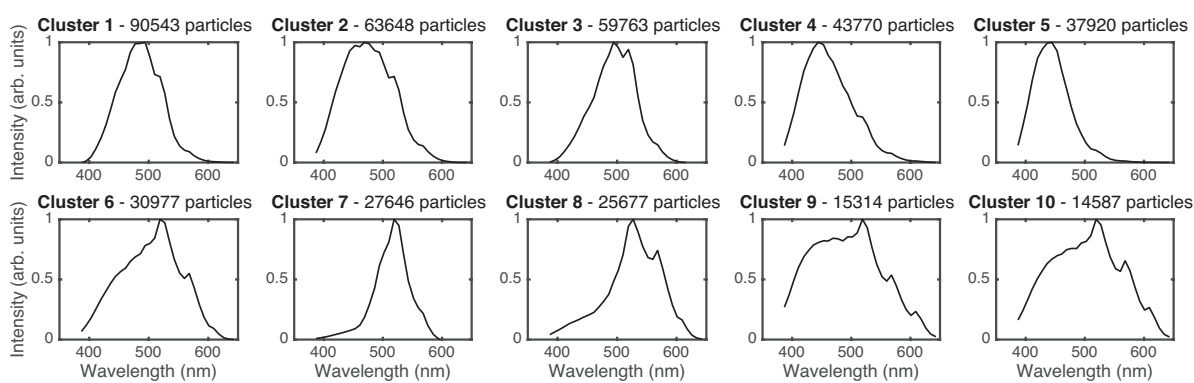

**Fig 2** The spectra of the 10 most significant clusters.

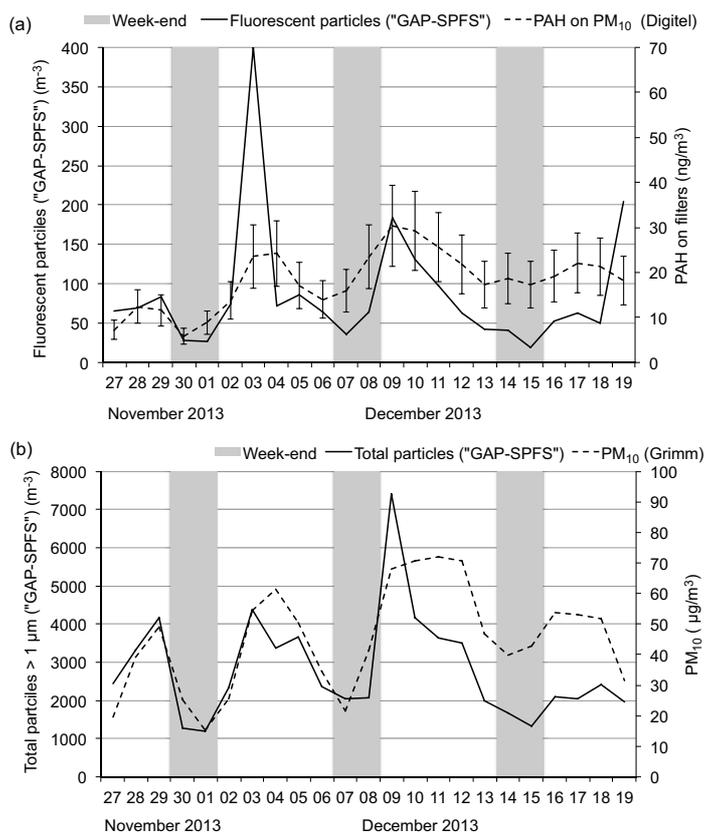

**Fig. 3** Comparison of the single-particle fluorescence spectrometer "GAP-SPFS" measurements with (a) total adsorbed PAH concentration on $PM_{10}$ from Digitel filters and (b) $PM_{10}$ concentration from a Grimm EDM #180.

To assess the validity of the GAP-SPFS measurements, we compared the day-to-day evolution of the PAH-bearing particles with that obtained from filter samples (Fig. 3a). In spite of completely different sampling methods and rates, the evolutions were parallel, with simultaneous maxima and minima still their respective amplitudes were different, so that the correlation coefficient between the two curves was 0.41 only. The amplitude discrepancy may be related to the different particle size ranges detected by both instruments. Furthermore, the filters yielded dry diameters while the SPFS measured the particle on the fly without drying. As expected for an urban site where PAH emissions are dominant, the fluorescent particles accounted most of the time for 5 to 10% of all detected particles, with peaks of up to 20% (**Fig. 4**b).

Similarly, the total particle concentration detected by the GAP-SPFS was correlated with the $PM_{10}$ ($r = 0.64$), as measured by the Grimm EDM #180 standard optical particle monitor / analyzer (Fig. 3b). Both devices measured the same decay in the aerosol load during the week-ends, due to a lower traffic. Again, remaining discrepancies may be attributed to the differences in the size sensitivity ranges of the two instruments.

Unlike the filter sampling and GC-MS analysis of the PAH, the single-particle fluorescence spectrometer offered a virtually infinite temporal resolution as it measured and identified each particle individually in real-time, hence delivering the intra-day dynamics of the fluorescent particles. Fig. 4 displays the corresponding hourly-averaged values. The results gave access to the dynamics of the aerosols, which was governed by the particle emission, dispersion, and chemical evolution including aging. The latter refers to the chemical evolution of the organic-bearing particles, mostly governed by (photo)oxidation (Kim et al. 2009). In particular, the average day (Fig. 5) displayed more PAH-bearing particle at high traffic emission times. The morning and evening traffic peaks were clearly visible on week days. In contrast, during week-ends, where traffic was lower, the concentration stayed flat over the entire day. Such intra-day temporal dynamics could not be observed with filter measurements that required 24 h accumulation.

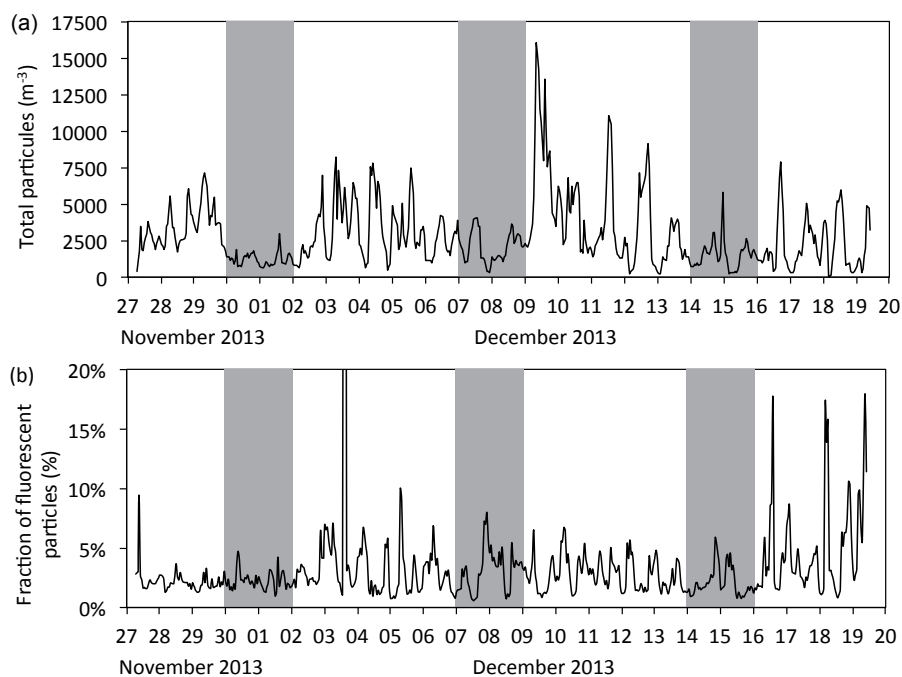

**Fig. 4** (a) Total particle number and (b) fraction of fluorescent particles, as measured by the GAP-SPFS. Grey bands denote week-ends

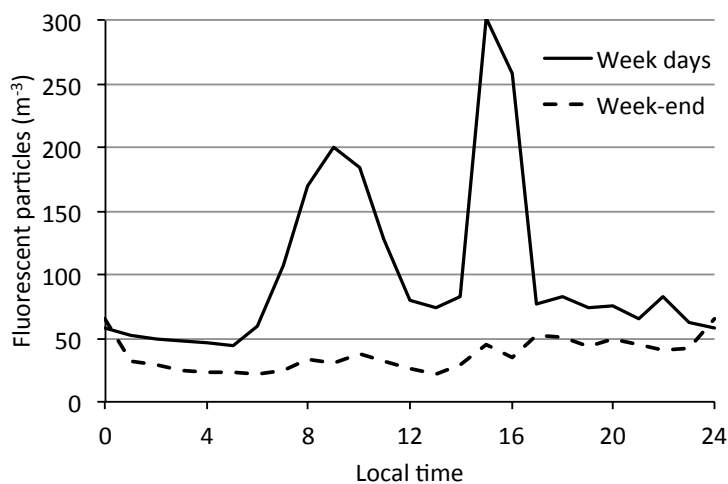

**Fig. 5** Average intra-day evolution of the concentration of fluorescent (PAH-bearing) particles. NB: Figures will be polished once we validate the contents

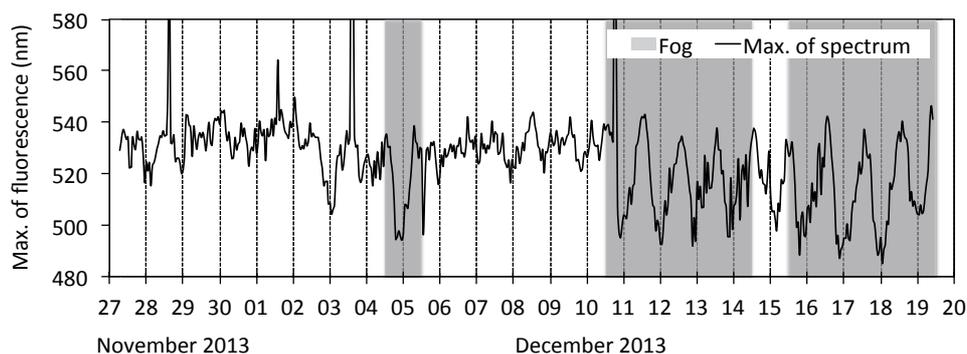

**Fig. 6** Evolution of the peak wavelength of the fluorescence spectrum and fog episodes.

The high temporal resolution of the GAP-SPFS also evidenced a specific behavior of the fluorescence spectrum during fog episodes, where the average peak fluorescence wavelength was blue-shifted by more than 40 nm each night (Fig. 6), and red-shifts back at day. This oscillation widely exceeds the standard deviation associated to this average, that is slightly below 20 nm. Such red-shift in humid conditions in the presence of ozone has indeed been observed by Pan et al. (2014) due to selective oxidation of peptides, especially tryptophan into kyunerenine. Simultaneously, the relative abundance of particles larger than 1 µm decayed or kept low (Fig. 4a). The nighttime frequency shift could therefore also be influenced by the maximum emission of biomass burning associated with heating (Mohr et al. 2013; Healy et al., 2012).

Indeed, the uptake of organics into fog particles (Birdwell and Valsaraj, 2010) can be expected to concentrate these species as well as to allow aqueous-phase reactions to occur (Herckes et al. 2013), increasing the rate of reactions leading to the formation of brown carbon (Laskin et al. 2015), with time constants of a few hours or less (Lee et al. 2014). At day, this brown carbon may undergo aging, both in and outside fog particles. Such aging of brown carbon and its constituents has a time constant of several hours (Kim et al. 2009) comparable to the sedimentation time of micrometer-sized particles (Seinfeld and Pandis 2006) and is generally associated with a red-shift of the fluorescence spectrum (Chang and Thompson, 2010; Rincón et al. 2009). It could therefore explain the observed red-shift of the particle fluorescent spectrum at daytime.

Although the above interpretations would need confirmation, such observation of the aging dynamics illustrates the potential of real-time fluorescence analysis to reveal fast and complex events at the particulate level.

## Conclusion

As a conclusion, a single-particle fluorescence spectrometer allowed to sort individual PAH-bearing particles in real time. The time-averaged results were consistent with classical particle detectors as well as optical particle analyser or filter measurements and subsequent analysis in the laboratory. Furthermore, the real-time resolution provided information on the particle physico-chemical dynamics, governed by their emission, dispersion and chemical evolution. The latter included aging of the atmospheric PAH adsorbed on aerosol particles. In particular, it offered a distinction between background emissions related to heating, and the traffic peaks during rush hours. It also evidenced the aging of the particles during the day.

Further work will be required to associate fluorescent clusters with specific sources of aerosols as well as to determine in detail their fate in the atmospheric compartment. Such detailed characterization could rely on the sampling and chemical analysis of individual particles, or on more advanced optical characterization currently under development (Sousa et al. 2016). It would refine the characterization of the adsorbed PAH, e.g., by allowing the measurement of their aging. It may therefore ultimately provide both a better understanding of the PAH dynamics in the atmosphere, and a more precise evaluation of population exposure under immission conditions.

## References

Birdwell, J. E., Valsaraj, K. T., (2010) Characterization of dissolved organic matter in fogwater by excitation–emission matrix fluorescence spectroscopy. *Atmospheric Environment*, *44*, 3246. Doi: 10.1016/j.atmosenv.2010.05.055

Bonacina, L., Kiselev, D., Wolf, J.P. (2013) Measurement Device and Method for Detection of Airborne Particles, European Patent EP 12167800.7 and US Patent 2013/0301047A1

Chang, J. L., Thompson, J. E.. (2010) Characterization of colored products formed during irradiation of aqueous solutions containing H2O2 and phenolic compounds, *Atmospheric Environment, 44*, 541–551. Doi: 10.1016/j.atmosenv.2009.10.042

Commission directive (EU)(2015) 2015/1480 of 28 August 2015 amending several annexes to Directives 2004/107/EC and 2008/50/EC of the European Parliament and of the Council laying down the rules concerning reference methods, data validation and location of sampling points for the assessment of ambient air quality. Official Journal of the European Union L 226/4 (29/8/2015): 4-11

EU-Commission (2004). Directive 2004/07/EC of the European parliament and the council of 15 December 2004 relating to arsenic, cadmium, mercury, nickel and polycyclic aromatic hydrocarbons in ambient air. *Official Journal of the European Communities* L 23(26/1/2005): 3-16

Finlayson-Pitts, B. J., Pitts, Jr, J. N, (2000) Chemistry of the Upper and Lower Atmosphere: Theory, Experiments, and applications, Academic Press, San Diego, pp 436–546

Hayashida, K., Amagai, K., Satoh, K., Arai, M., (2006) Experimental analysis of soot formation in sooting diffusion flame by using laser-induced emissions, *Journal of Engineering for gas turbines and power*, *128*, 241. doi:10.1115/1.2056536

Healy, R. M et al. (2012) Sources and mixing state of size-resolved elemental carbon particles in a European megacity: Paris. *Atmospheric Chemistry and Physics*, *12*, 1681–1700. doi: 10.5194/acp-12-1681-2012

Herckes, P., Valsaraj, K. T., Collett Jr., J. L., (2013) A review of observations of organic matter in fogs and clouds: Origin, processing and fate, *Atmospheric Research*, *132–133*, 434–449. Doi: 10.1016/j.atmosres.2013.06.005

Herrmann, H., Brüggemann, E., Franck, U., Gnauk, T., Löschau, G., Müller, K., Plewka, A., Spindler, G. (2006) A source study of PM in Saxony by size-segregated characterisation, *Journal of Atmospheric Chemistry, 55*, 103–130

Hill, S.C., et al. (1999) Real-Time Measurement of Fluorescence Spectra from Single Airborne Biological Particles, Field analytical chemistry and technology 3:221–239. DOI: 10.1002/(SICI)1520-6521(1999)3:4/5<221::AID-FACT2>3.0.CO;2-7

Huffman, J. A., *et al.*, (2013), High concentrations of biological aerosol particles and ice nuclei during and after rain, *Atmos. Chem. Phys. 13*, 6151–6164.

Ito K., Kinney P.L., Thurston G.D. (1995) Variations in $PM_{10}$ concentrations within two metropolitan areas and their implications for health effects analyses. *Inhal Toxicol*, 7, 735–745. Doi: 10.3109/08958379509014477

Kasparian et al. (2017) Assessing the Dynamics of Organic Aerosols over the North Atlantic Ocean, *Scientific Report* submitted.

Kim, D., *et al.* (2009) Environmental aging of polycyclic aromatic hydrocarbons on soot and its effect on source identification, *Chemosphere*. *76,* 1075–1081. doi: 10.1016/j.chemosphere.2009.04.031

Kiselev, D., Bonacina, L., Wolf, J.P., (2013) A flash-lamp based device for fluorescence detection and identification of individual pollen grains, *Rev. Sci. Instrum*. 84: 033302. DOI: 10.1063/1.4793792

Laskin, A., Laskin, J., Nizkorodov, S. A. (2015) Chemistry of Atmospheric Brown Carbon, *Chemical Reviews*, *115*, 4335. DOI: 10.1021/cr5006167

Lee, H. J. (Julie), Aiona, P. K., Laskin, A., Laskin, J., Nizkorodov, S. A., (2014) Effect of Solar Radiation on the Optical Properties and Molecular Composition of Laboratory Proxies of Atmospheric Brown Carbon, *Environ. Sci. Technol*, *48*, 10217–10226. DOI: 10.1021/es502515r

Maki, T., *et al.*, (2008), Phylogenetic diversity and vertical distribution of a halobacterial community in the atmosphere of an Asian dust (KOSA) source region, Dunhuang City *Air Qual. Atmos. Health, 1*, 81–89

Miyakawa, T., *et al.*, (2015) Ground-based measurement of fluorescent aerosol particles in Tokyo in the spring of 2013: Potential impacts of nonbiological materials on autofluorescence measurements of airborne particles *J. Geophys. Res. Atmospheres*, *120*, 1171–1185. DOI: 10.1002/2014JD022189

Mohr, C., *et al.*, (2013) Contrubtion of nitrated phenols to wood burning brown carbon light absorption in Detling, United Kingdom during winter time, *Environmental Science and Technology*, *47*, 6316. DOI: 10.1021/es400683v

Pan, Y., et al., (2001) High-speed, High-sensitivity Aerosol Fluorescence Spectrum Detection Using 32-Anode PMT Detector, Rev.Sci.Inst. 72:1831-1836. DOI: 10.1063/1.1344179

Pan, Y.L., *et al.*, (2003) Single-Particle Fluorescence Spectrometer for Ambient Aerosols, *Aerosol Science and Technology*, *37*, 628–639. Doi: 10.1080/02786820300904

Pan, Y.-L., Huang, H., Chang, R. K., (2012) Clustered and integrated fluorescence spectra from single atmospheric aerosol particles excited by a 263- and 351-nm laser at New Haven, CT, and Adelphi, MD, *Journal of Quantitative Spectroscopy and Radiative Transfer*, *113*, 2213. Doi: 10.1016/j.jqsrt.2012.07.028

Pan Y-L., Santarpia J., Ratnesar-Shumate S., Corson E., Eshbaugh J., Hill S., Williamson C., Coleman M., Bare C., Kinahan S., (2014) Effects of ozone and relative humidity on fluorescence spectra of octapeptide bioaerosol particles. Journal of Quantitative Spectroscopy and Radiative Transfer 133:538-550. DOI: 10.1016/j.jqsrt.2013.09.017

Pan, Y.L., (2015) Detection and characterization of biological and other organic-carbon aerosol particles in atmosphere using fluorescence, *J. Quant. Spec. and Rad. Trans.*, *150*, 12–35. Doi: 10.1016/j.jqsrt.2014.06.007

Phillips, S. M., Smith, G. D. (2015) Further Evicence for Charge Transfer Complexes in Brown Carbon Aerosols from Excitation-Emission Matrix Fluorescence Spectroscopy, *The Journal of Physical Chemistry A*, *119*, 4545-4551. DOI: 10.1021/jp510709e

Pinnick, R. G., *et al.*, (2013) Fluorescence spectra and elastic scattering characteristics of atmospheric aerosol in Las Cruces, New Mexico, USA: Variability of concentrations and possible constituents and sources of particles in various spectral clusters *Atmos. Environment*, *65*, 195-204. Doi: 10.1016/j.atmosenv.2012.09.020

Pinnick, R. G., *et al.* (2004) Fluorescence spectra of atmospheric aerosol at Adelphi, Maryland, USA: measurement and classification of single particles containing organic carbon *Atmospheric Environment*, *38* 1657–1672. Doi: 10.1016/j.atmosenv.2003.11.017

Pöhlker, C., Huffman, J. A., Pöschl, U. (2012) Autofluorescence of atmospheric bioaerosols – fluorescent biomolecules and potential interferences. *Atmos. Meas. Tech.*, *5*, 37–71. Doi: doi: 10.5194/amt-5-37-2012

Pope, C. A. and Dockery, D. W. (2006) Health Effects of Fine Particulate Air Pollution: Lines that Connect *J. Air Waste Manage. Assoc.*, *56*, 709–742. Doi: 10.1080/10473289.2006.10464485

Rincón, A. G., Guzmán, M. I., Hoffmann, M. R., Colussi, A. J. (2009) Optical Absorptivity versus Molecular Composition of Model Organic Aerosol Matter, *The Journal of Physical Chemistry A*, *113*, 10512–10520. DOI: 10.1021/jp904644n

Robinson, N. H., *et al.*, (2013) Cluster analysis of WIBS single-particle bioaerosol data, *Atmos. Meas. Tech.*, *6*, 337–347. doi:10.5194/amt-6-337-2013

Seinfeld, J and Pandis, S (2006) Atmospheric Chemistry and Physics: From Air Pollution to Climate Change. John Wiley & Sons

Saari, S.E., Putkiranta, M. J., Keskinen, J. (2013) Fluorescence spectroscopy of atmospherically relevant bacterial and fungal spores and potential interferences, *Atmospheric Environment 71*, 202–209

Sousa, G., Gaulier, G., Bonacina, L., Wolf, J.-P. (2016) Portable Instrument Discriminating Bio-aerosols from Non-Bio-aerosols using Pump-Probe Spectroscopy, *Scientific Report* 6, 33157. doi:10.1038/srep33157